\documentclass[aps,pre,twocolumn,superscriptaddress]{revtex4-2}
\usepackage{amssymb}
\usepackage{amsmath,bm}
\usepackage{graphicx,color}
\usepackage{bbm}
\usepackage[bottom]{footmisc}
\usepackage{multirow}
\usepackage{xcolor}
\usepackage{xpatch} 
\makeatletter   
\usepackage{xpatch} 

\newcommand{\B}[1]{{\bm{#1}}}

\newcommand{\rin}{r_\text{in}}
\newcommand{\rout}{r_\text{out}}
\newcommand{\C}[1]{{\mathcal{#1}}}

\begin{document}
\title{Dipole-Induced Transition in 3-Dimensions}
\author{Itamar Procaccia} 
\affiliation{Dept. of Chemical Physics, The Weizmann Institute of Science, Rehovot 76100, Israel}
\affiliation{Sino-Europe Complex Science Center, School of Mathematics, North University of China, Shanxi, Taiyuan 030051, China.}
\author{Tuhin Samanta}
\affiliation{Dept. of Chemical Physics, The Weizmann Institute of Science, Rehovot 76100, Israel}

\begin{abstract}
The Kosterlitz-Thouless and the Hexatic phase transitions are celebrated examples of dipole (vortex, dislocation) induced transitions in condensed matter physics. For very clear reasons, these important ``topological" transitions are restricted to 2-dimensions. Here we present a genuine dipole-induced transition in the 3-dimensional response of (athermal) amorphous solids to applied strain. Similarly to the existence of a hexatic phase between normal solid and fluid, we identify an intermediate phase between a phase of normal elastic response at high pressure, and fluid matter at zero pressure. The mechanical response in the intermediate phase is accompanied by plasticity that is generically associated with ``non-affine" quadrupolar events seen in the resulting displacement field. Gradients of the quadrupolar fields act as dipole charges that screen elasticity, breaking both translational and Chiral symmetries. We highlight {\em angular} correlations that exhibit diverging correlation lengths at this transition and determine the critical scaling exponents.
\end{abstract}
\maketitle


{\bf Introduction}:  The celebrated Hohenberg-Mermin-Wagner theorem in Statistical Mechanics states that in space dimension $d\le 2$ and in the thermodynamic limit, there is no phase with the spontaneous breaking of a continuous symmetry for any temperature $T > 0$ \cite{67Hoh,66MW}. On the other hand, this theorem does not mean that phase transitions cannot occur in 2-dimensions. The spontaneous nucleation of topological entities, like vortices, dislocations or dipoles can induce phase transitions of the Kosterlitz-Thouless type, in which a characteristic correlation length $\xi$ can diverge \cite{16Kos}. Implications for the $x-y$ model, as well as for superfluid and superconductor films, are well known and well studied \cite{18KT}. Another example of such a transition is the Hexatic transition that identifies an intermediate phase between long-ranged ordered 2-dimensional solid and liquid, in which hexatic order exists but long-range order is destroyed \cite{79NH}. On the other hand, this interesting type of ``topological transitions" are not known to occur in 3-dimensions. The intuitive reason for this is very clear: the energy cost of a dipole (or vortex, or dislocation) in 2-dimensions is logarithmic in the system size, but so is also the entropy, allowing the existence of a critical temperature where dipoles are free to proliferate. This balance between energy and entropy does not exist in 3 dimensions, limiting this type of thermal transition to 2-dimensions.

The aim of this Letter is to present and investigate a dipole-induced {\bf athermal} transition in three dimensions. Contrary to the thermal examples, the transition under study is not in the structure of the material or in the ordering of its degrees of freedom, but rather in the mechanical response of the material, which is an amorphous solid, to external strain. The parameter dictating the transition is the pressure, while the temperature
can remain zero (or sufficiently low) throughout. At large pressures, the amorphous solids studied below respond to strain quasi-elastically, even though there may exist some plastic events which are identified as quadrupoles in the ensuing displacement field. At zero pressure the material responds like a fluid. In analogy to the hexatic phase transition, we identify an intermediate phase at intermediate pressures for which dipoles in the displacement field dominate the response. Here characteristic screening lengths emerge spontaneously, leading to anomalous elastic responses. We will argue that near the transition between the intermediate and the high-pressure phases angular correlations appear to exhibit divergent correlation lengths, and we determine the associated scaling exponents.

{\bf System Details:} 
To set up the discussion we choose to exemplify the transition using an amorphous solid created {\it in silico} from $N$ frictionless small spheres of unit mass, prepared with a pressure $p$ in mechanical equilibrium. To avoid crystallization we employ bi-disperse spheres, randomly distributed within a spherical volume of outer radius $r_{\text{out}} \approx 22$, measured in terms of the smaller sphere's radius.  One-half of the spheres have a radius of $R_i= 0.5$, while the other half has a radius of $R_i= 0.7$, expressed in dimensionless units. One of the smaller spheres is fixed at the center of the sphere, and the studied configuration is bounded by a fixed outer spherical wall. The little spheres interact via
a normal Hertzian force. Denoting their center of mass coordinates as $\B r_i$, we define $\Delta_{ij} \equiv R_i+R_j -  |\B r_i -\B r_j|$. In equilibrium the inter-sphere normal force is $F_{ij} \equiv k_n \Delta_{ij}$, and $k_n =2000 \sqrt{\Delta_{ij} (R_i^{-1} +R_j^{-1})}$.
The simulation begins with a specific volume fraction corresponding to a desired equilibrium pressure, relaxed to mechanical equilibrium by solving Newton's second law of motion with damping applied. This process is continued until the target pressure is reached and the resultant force on each particle reaches values smaller than $10^{-6}$. All the quantities reported below are dimensionless. 

After equilibration, the central smaller sphere is inflated by $30\%$ and the response of the system is observed. We measure the displacement field, which is determined by subtracting the pre-inflation coordinate from the post-inflation one. This displacement field is where all the novel phenomena discussed below are hidden. 

\begin{figure} [h]
	\includegraphics[width=0.9\linewidth]{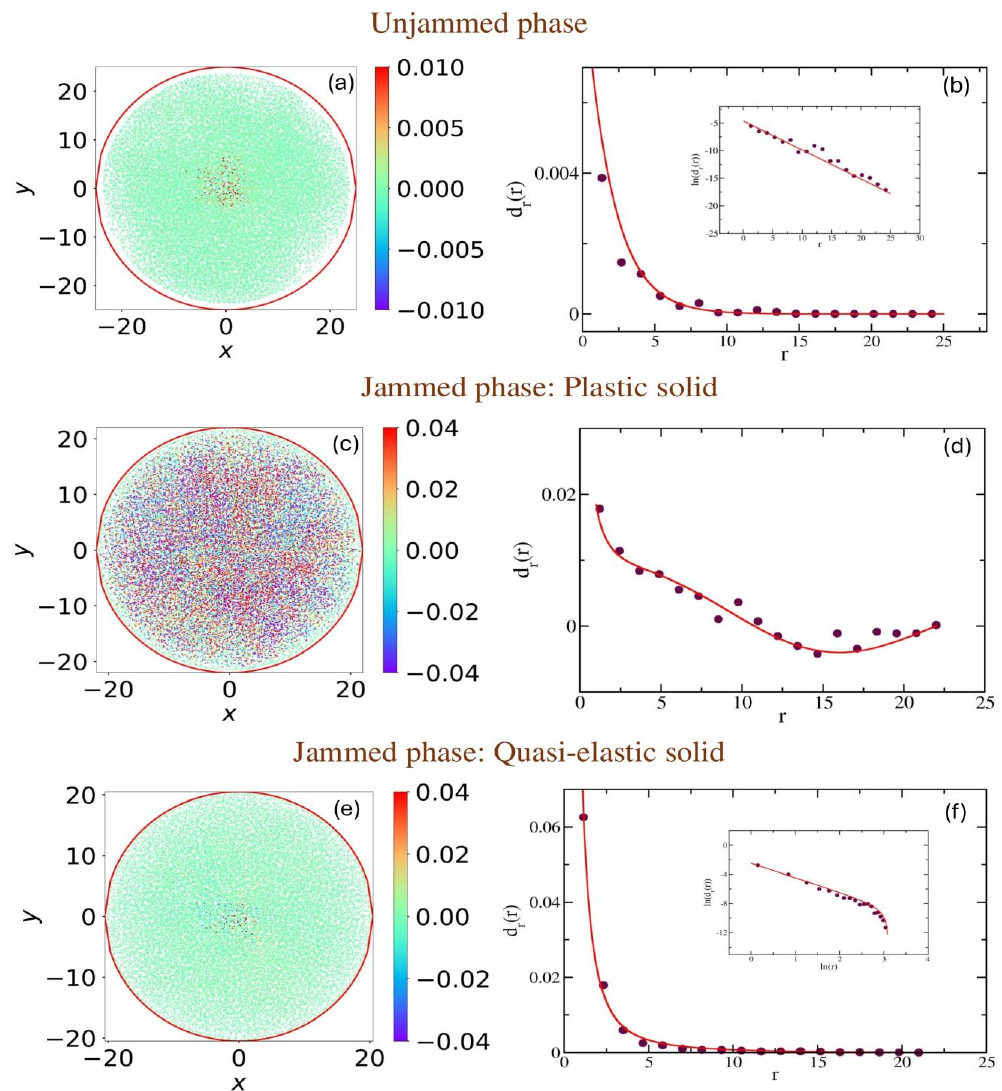}
	\caption{Panels (a) and (b): Map of the magnitude of $d_r(r)$ and a linear-linear plot of $d_r(r)$ vs. $r$ in the zero pressure phase, showing the exponential decay. The inset in panel (b) is the log-linear plot. Panels (c) and (d): The same as in panels (a) and (b) but in the intermediate phase of anomalous elasticity at $p=0.2$. The solid red line in panel (d) represents the solution Eq.~(\ref{amazing}) with $\kappa=0.285$. Panels (e) and (f): The same but in the high-pressure phase. The linear-linear plot in panel (f) shows the power-law decay of $d_r(r)$, with a log-log plot in the inset. The solid red line represents the solution of Eq.~(\ref{renelas}).}
	\label{results}
\end{figure}	

{\bf Typical displacement field for three different phases:}
Having the displacement field $\B d (r, \theta, \phi) $ in spherical coordinates, we extract the radial component $d_r(r)$ by angle averaging over the angles $\theta$ and $\phi$:
\begin{equation}
	d_r (r)\equiv \frac{1}{4\pi}\int_0^{2\pi} \sin \theta d\theta \int_0^\pi d\phi d_r(r,\theta,\phi) \ .
\end{equation}
Typical results are shown in Fig.~\ref{results}. At zero pressure, (upper panels (a) and (b)), when the packing fraction is set below the jamming transition $\Phi_j\approx 0.64$, the responding radial component of the displacement field decays exponentially with a characteristic correlation length $\zeta$. At high pressures, (panels (e) and (f)), when the packing fraction is much larger than $\Phi_j$, the response is quasi-elastic, with the radial component decaying as a power-law in the bulk, with a characteristic $r$-dependence going like $r^{-2}$. The most interesting case is that of the intermediate value of the packing fraction, where one finds a response that is neither exponential nor a power law. As can be seen from panels (c) and (d), although the inflation at the center points outward, the radial component of the displacement changes sign, with the particle moving inward rather than outward. We will see below that this behavior stems from dipole screening that introduces characteristic screening lengths that determine the anomalous form of the displacement field.

{\bf Theoretical considerations:}
The understanding of the different responses shown in Fig.~\ref{results} is based on a theory that was developed over the last few years \cite{21LMMPRS,22BMP,23CMP}, a theory that was demonstrated in a variety of experimental \cite{22MMPRSZ} and simulational contexts \cite{22KMPS,23MMPR,24HPPS}. In short, 
for a classical elastic solid the equation obeyed by displacement field $\B {d}$ reads
\begin{equation}\label{L1}
	\mu \Delta \B {d} + (\lambda +\mu)\B \nabla (\B \nabla \cdot \B{d}) =0.
\end{equation}
Here $\lambda$ and $\mu$ are the classical Lam\'e coefficients. This equation suffices to describe the power-law response that we find in the high-pressure phase, cf. panels (e) and (f) in Fig.~\ref{results}. Solving this equation with the boundary conditions $\d(r_{\rm in}, \theta, \phi) =d_0$ and  $\d(r_{\rm out}, \theta, \phi) =0$, we find 
\begin{equation}
	d_r (r)= d_0 \frac{r_{\text{in}}^2 \left(r^3-r_{\text{out}}^3\right)}{r^2 \left(r_{\text{in}}^3-r_{\text{out}}^3\right)}\ .
	\label{renelas}
\end{equation}
In the bulk of the system, when $\rin< r<\rout$, the solution decays as $1/r^2$, as expected in standard elasticity theory. This is the behavior seen in panel (f) of Fig.~\ref{results}.

In the intermediate phase, things are different. First, generically any strain on amorphous solids results in plastic deformation \cite{10KLP,11HKLP}. Typical plastic events are quadrupolar in nature, called sometimes ``Eshelby inclusions" due to the resemblance to the Eshelby solution of the displacement field caused by forcing a circle to change to an ellipse within the material \cite{54Esh}. When the density of quadrupolar events is finite, one refers to the resulting quadrupolar field as $Q^{\alpha\beta}(\B r)$. It was shown that the creation of such a field leads to a renormalization of the elastic moduli \cite{21LMMPRS,23CMP}. Further, when the quadropolar {\em field} of such events is non-uniform, it was explained that gradients of the field act as effective dipoles,
\begin{equation}
	\C	P^\alpha \equiv \partial_\beta Q^{\alpha\beta} \ .
\end{equation}
When these are present, the equation satisfied by the displacement field changes in a qualitative way, reading \cite{21LMMPRS,23CMP}
\begin{equation}\label{L2}
	\mu \Delta \B {d} + (\lambda +\mu)\B \nabla (\B \nabla \cdot \B {d}) +\B \Gamma \B {d} =0.
\end{equation}
Here $\B \Gamma$ is a tensor that needs to be specified, containing screening parameters \cite{MM24odd, TS25}. In the simplest version of the theory one models the tensor
$\Gamma$ as diagonal, leading to an equation to be solved of the form  
\begin{equation}\label{L2}
	\mu \Delta \B {d} + (\lambda +\mu)\B \nabla (\B \nabla \cdot \B {d}) +k^2\B {d} =0.
\end{equation}
The term $k^2\B {d}$ is responsible for translational symmetry breaking, the introduction of a typical length scale $\ell$, 
$\ell \sim k^{-1}$, and to screening phenomena that change dramatically the expected displacement field $\B d$ from the predictions of Eq.~(\ref{L1}). We note in passing that if one is interested in the tangential component of the displacement field, one needs to consider a more general form of the tensor $\B \Gamma$. Details of this and the associated Chiral symmetry breaking can be found in Refs.\cite{24LSSPM,TS25}. For our purposes here the simple model of diagonal tensor will suffice. Below we will focus on the radial component only, and on fluctuations of the radial component as a function of the angles $\theta$ and $\phi$.

With the same boundary conditions as before, the analytic solution of Eq.(\ref{L2}) reads \cite{21LMMPRS,23CMP}
\begin{equation}
	d_r(r)  = d_0 \frac{ Y_1(r \, \kappa ) J_1(r_\text{out} \kappa )-J_1(r \, \kappa ) Y_1(r_\text{out} \kappa )}{Y_1( r_\text{in} \kappa ) J_1(r_\text{out} \kappa )-J_1(r_\text{in} \kappa ) Y_1(r_\text{out} \kappa )} \ .
	\label{amazing}
\end{equation}
Here $J_1$ and $Y_1$ are the spherical Bessel functions of the first and second kind respectively. The continuous line in panel (d) that fits nicely the data points is precisely this solution with $\kappa\approx 0.285$.

\begin{figure} [h]
	\includegraphics[width=0.9\linewidth]{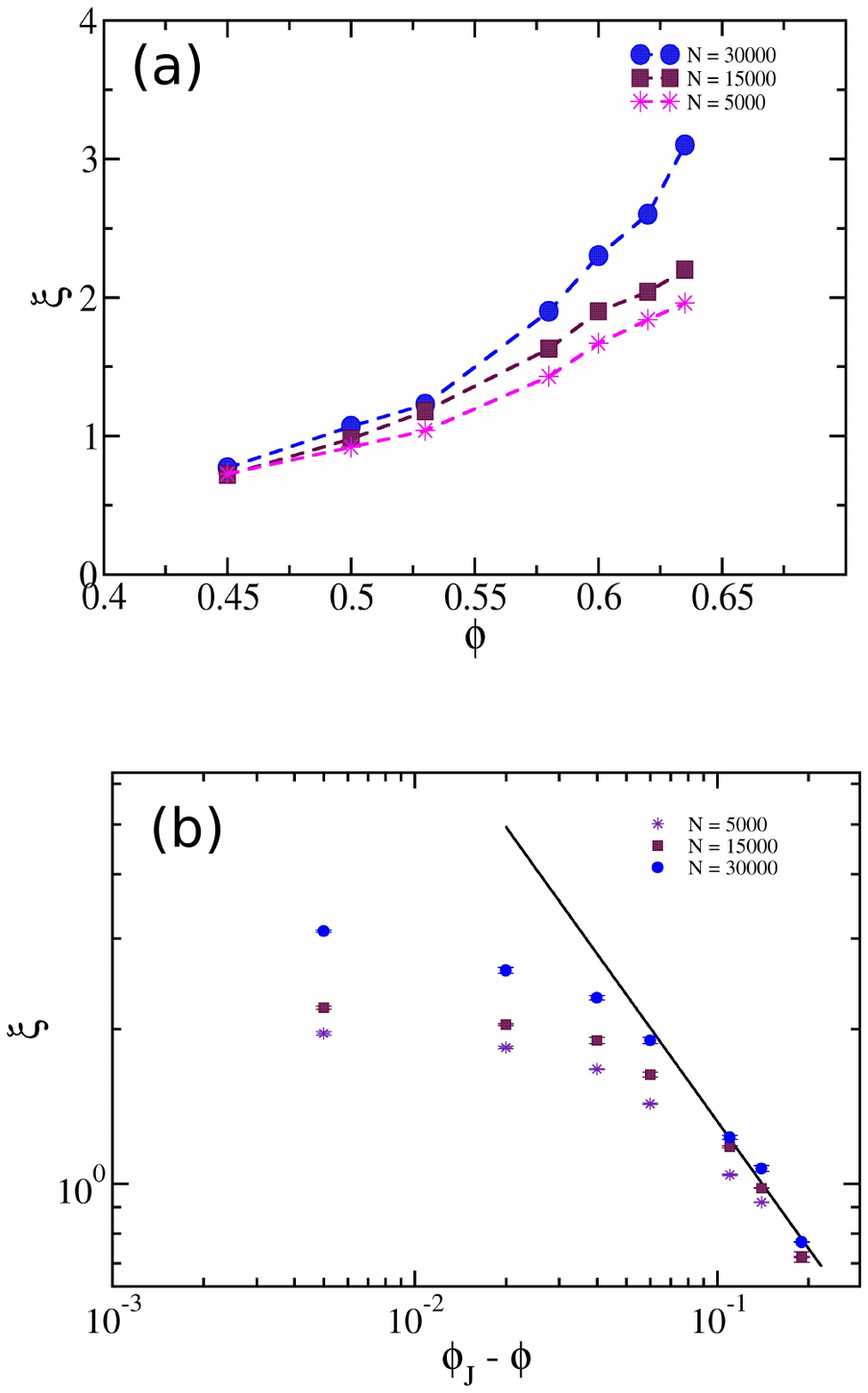}	
	\caption{Pnael a: The radial decay length of the displacement in the unjammed fluid phase as a function of the distance from jamming $\Phi_j\approx 0.64$. The divergence becomes more apparent for larger system sizes. Panel b: the same data in the log-log plot, to estimate the scaling exponent $\nu$. }
	\label{xi}
\end{figure}

{\bf Nature of the transitions between the three phases:}
The first (well-known) transition is the jamming transition which has been extensively investigated in a series of previous studies \cite{18BMHM,OH03,ML00}. The jamming density and the associated transitions are notably protocol-dependent and dependent on compression and preparation methods \cite{10CBS, FZ12,MV14}. In the un-jammed phase, the displacement field simply falls off exponentially with the distance from the inflating central sphere. The exponential functions have a typical length $\xi$ that diverges towards the jamming transition. This is well known, and in our simulations we simply tested that the expected divergence is found, with a power-law dependence on the volume fraction, $\xi\sim (\Phi_j-\Phi)^{-\nu}$, with $\Phi_j\approx 0.64$ and $\nu\approx 0.8$, see Fig.~\ref{xi}. This value of exponent is in close correspondence with well known results ~\cite{NL02,JL09,PAN2023,PT11,OH03,Abram19,HN18,fu24}. 

\begin{figure} [h]
	\includegraphics[width=0.9\linewidth]{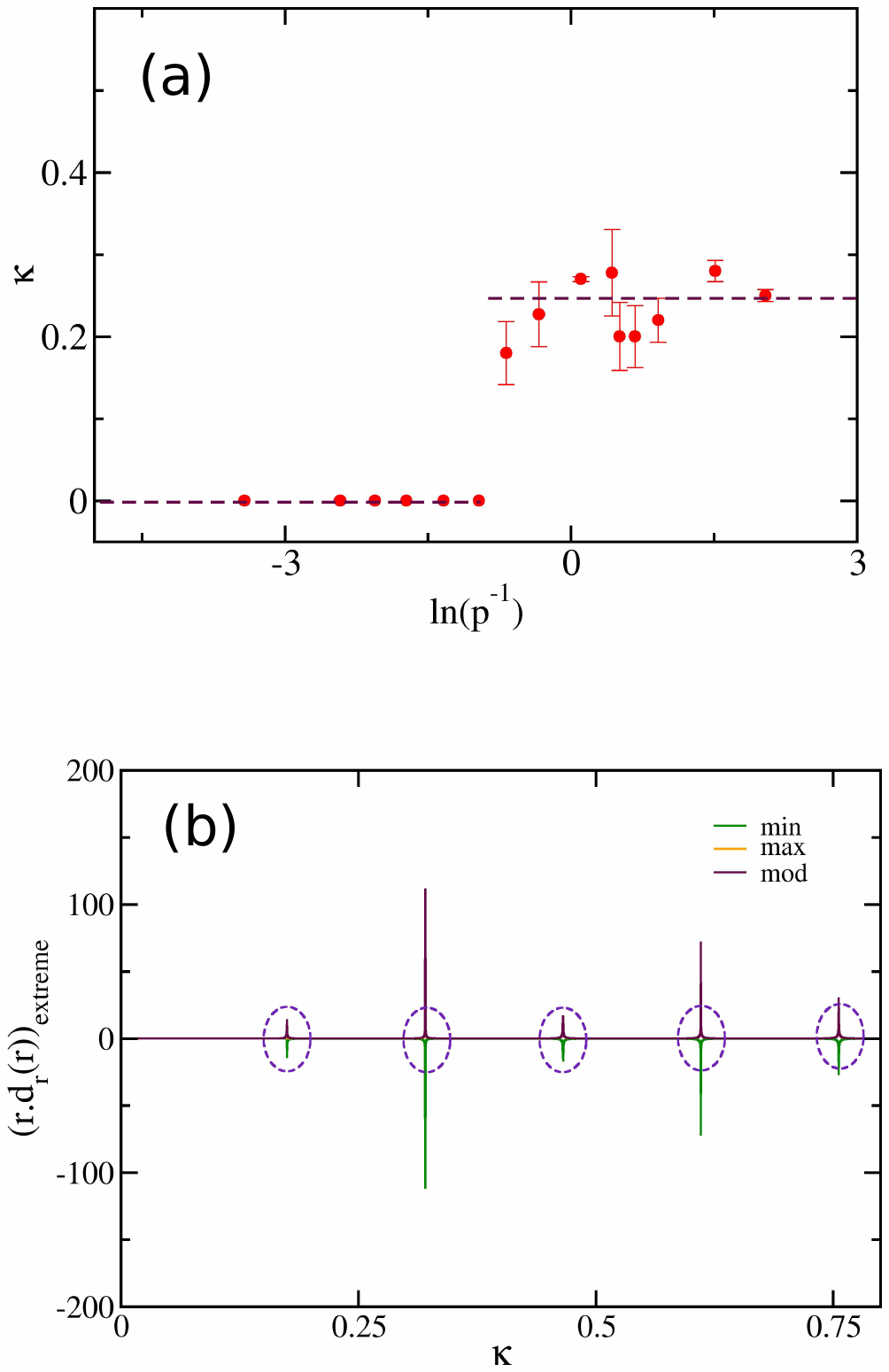}
	\caption{Panel a: the jump in the observed value of the screening parameter $\kappa$ (cf. Eq.~(\ref{amazing})) as the pressure reduces towards $p_c$. 
		The selected value of $\kappa\approx 0.256$ is well understood, it is determined by the first admissible singularity in the maximal amplitude of the solution (\ref{amazing}), (value of $\kappa\approx 0.29$ for which the denominator goes to zero. see panel (b)). Panel (b): The maximal amplitude of $r d_r(r)$ as computed from Eq.~(\ref{amazing}). We see the preferred values of $\kappa$ at which the anomalous response is maximal. The first singularity requires a $\kappa$ that is too small, or $\ell$ too large for our system size, and thus the second singularity determines the observed value of the screening parameter.}
	\label{sharp}
\end{figure}

The second transition \cite{21LMMPRS,23CMP,24JPS} is between the quasi-elastic phase at high pressures and an intermediate phase that exists at pressures $p_c\ge p \ge 0$, where $p_c$ is the critical pressure below which dipole induced screening is observed. The transition can be identified by the jump in the screening parameter $\kappa$ from zero to a finite value at $p_c\approx 2.3 \pm 0.3 $, see panel (a) of Fig.~\ref{sharp}. 
One should note that the determination of $p_c$ from these data is only approximate, and we will see additional estimates below. In addition, it should be stated that $p_c$ depends on the amount $d_0$ of inflation and on the geometric parameters $r_{\rm in}$ and $r_{\rm out}$.

As explained in detail in Refs.~\cite{24JPS} and \cite{TS25}, the emergent value of the screening parameter is theoretically predictable; it has to do with one of the discrete values of $\kappa$ that renders the solution (\ref{amazing}) singular as seen in panel (b) of Fig.~\ref{sharp}. The first two theoretical predicted values of $\kappa$ ($\kappa \approx 0.145$ and $\kappa \approx 0.29$) are illustrated in panel (b) of Fig.~\ref{sharp}. In the present case it is the second singularity that is important - the first one would select a value of $\kappa$ that is too small for our system size, as explained below. The point $p \approx p_c$ where the solution jumps from quasi-elastic to anomalous is the transition that we want to highlight. Due to finite size effects, including apparent sample-to-sample fluctuations in $\kappa$ within the anomalous region, the precise determination of the transition point will be made with a different approach, which uses an important characteristic of the transition, which is the gain (or loss) of angular correlations. 

\begin{figure} [h]
	\includegraphics[width=1.0\linewidth]{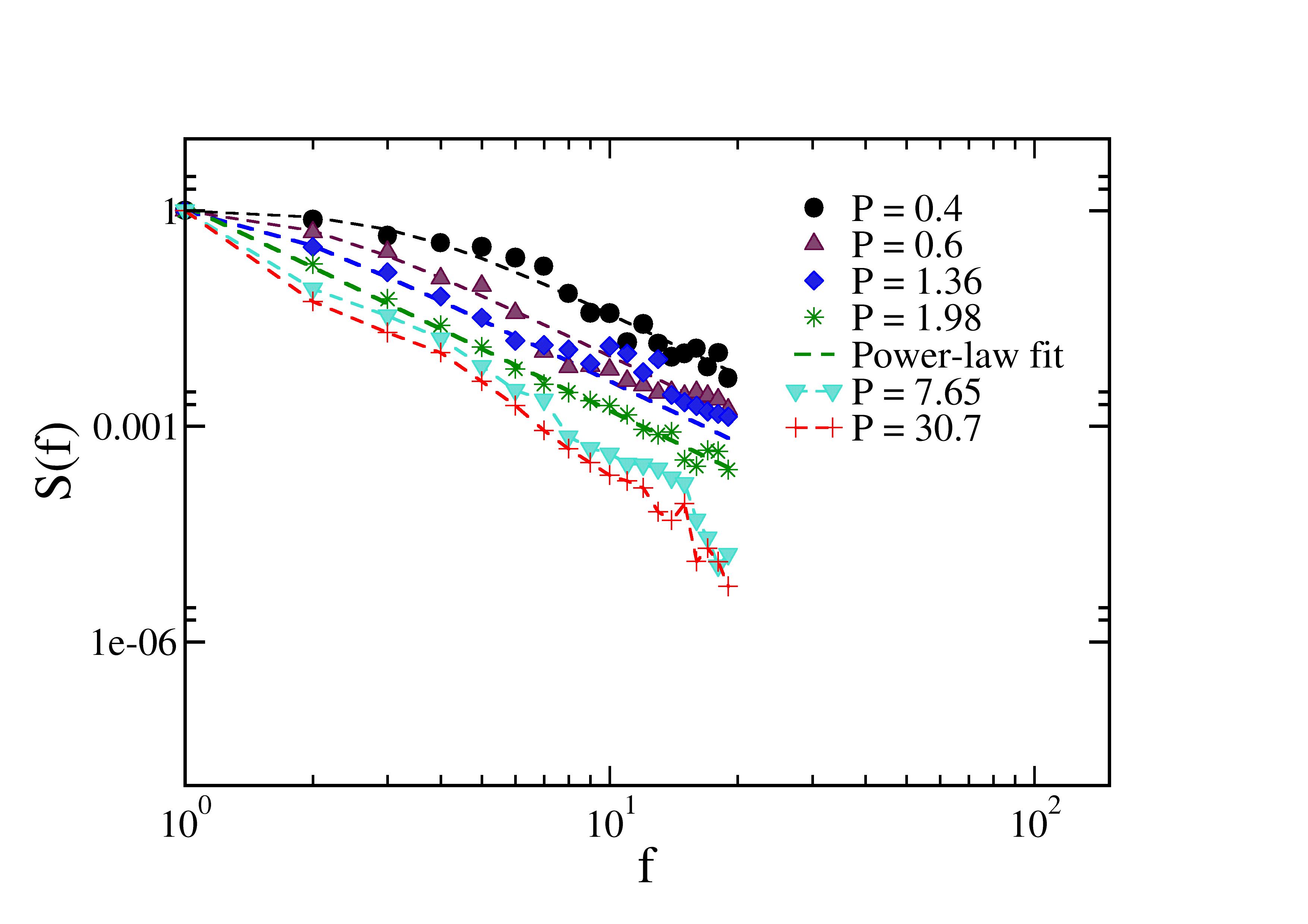}
	\caption{The power spectrum $S(r,f)$ for different values of the pressure. }
	\label{Sf}
\end{figure}

To understand this crucial point, recall that in the quasi-elastic regime, $p> p_c$, the system still satisfies the classical equation 
(\ref{L1}). The solution for the radial component of the displacement field $d_r(r,\theta, \phi)$ is expressed as a complete set of functions which for the sphere are the Legendre polynomials, (instead of sines and cosines in 2-dimensions). This means that in this phase, when we travel along any circle of radius $r$, angular correlations exist throughout. This is no longer correct in the anomalous phase. To quantify and employ this fact, we will consider the Fourier transform the displacement field according to

\begin{equation}
X(r, \phi, f) \equiv \frac{1}{M} \sum_{m=1}^M  d_r(r_m, \theta_m, \phi) \exp\left(-i  f \theta_m\right),
\end{equation}
where the sum runs on all the $M$ particles whose center of mass resides in an annular ring of radius
$r$ and width $\delta r$, with M  angular coordinates $\theta_m$ and a fixed $\phi$. The resulting function is averaged over independent configurations prepared with the same conditions. Next we consider the function $S(r,f)$,
\begin{eqnarray}
	S(r,f) \equiv |\langle 	X(r,\phi,f)\rangle |^2 \ ,
\end{eqnarray}
where pointed brackets stand for ensemble average. Due to this averaging the resulting function $S(r,f)$ loses the dependence on $\phi$. In Fig.~\ref{Sf}
we present typical results for this function at $r= 7$ and $\delta r \approx 2.5$.
We draw the reader's attention to the change in the functional form of $S(r,f)$ for the pressure values $p<p_c$, tending towards a power-law fit when the pressure reaches the critical value $p\approx p_c$. This should supply a precise estimate of the transition pressure, as we show next. 

\begin{figure} [h]
	\includegraphics[width=0.8\linewidth]{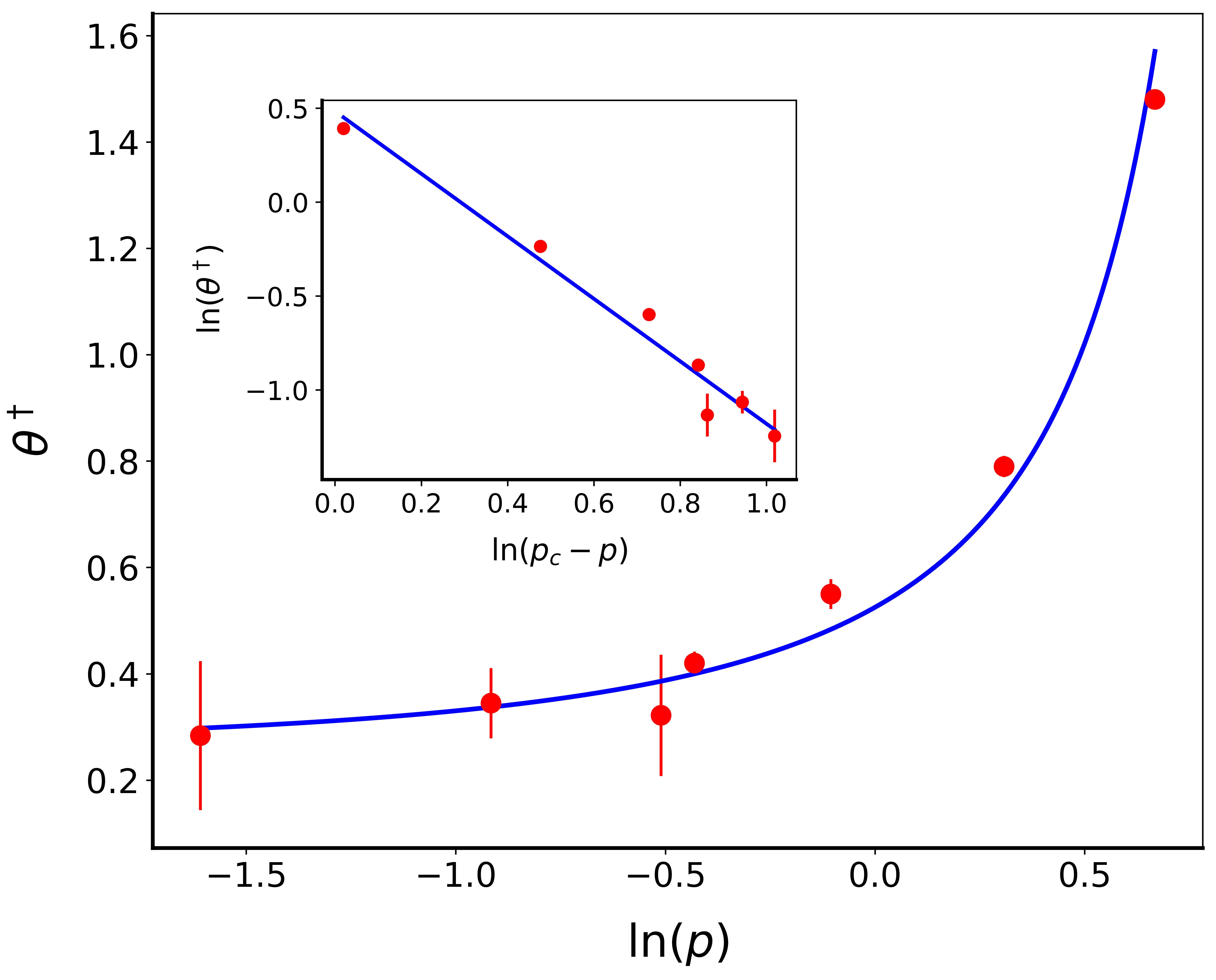}
	\caption{The angular correlation length $\theta^\dag$ as a function of pressure.
	Inset, log-log plot of the $\theta^\dag$ as a function of $p_c-p$ with $p_c \approx 2.9$. }
	\label{thetadag}
\end{figure}

The dashed lines that go through the data points in Fig.~\ref{Sf} at pressures below $p_c$ are fits to a generalized Lorentzian form \cite{toda2012}
\begin{equation}
S_{GL}(r,f) \equiv \frac {A}{1+(f/f^\dag)^{2.85}} \ ,
\end{equation}
where $A=1+(1/f^\dag)^{2.85}$ is a normalization factor and $f^\dag$ is the only fit parameter. Clearly, as the power spectrum approaches a pure power law, $f^\dag \to 0$. Accordingly, we define a correlation length $\theta^\dag\equiv 1/f^\dag$, and plot this quantity as a function of pressure. The results are shown in Fig.~\ref{thetadag}.
The power law presented in the inset is the main result of this paper, indicating an apparent critical divergence of the angular correlation length
\begin{equation}
\theta^\dag \propto (p_c-p)^{-\mu} \ , \quad \mu \approx 1.66 \quad p_c\approx 2.9 \ .
\label{result}
\end{equation}

{\bf Theoretical estimate of the transition pressure $p_c$:}
Finally, we need to understand what selects the critical pressure $p_c$.  To understand at which pressure the anomalous solution would appear, 
we need to determine when we can expect an avalanche of plastic events that can span a region of size $\kappa^{-1}$. 

Plastic avalanches start typically with a localized quadrupolar events, and the question is whether this even will cause and avalanche of plastic events, and if it does, of what size. At the jamming point, it is sufficient for a plastic event to cut one contact between particles to lose the rigidity of the whole system – the whole system collapses as a rigid solid and becomes liquid. As the pressure increases, when the average number of contacts exceeds more and more the critical value of 6, a plastic event can result in cutting more than one bond, but only a region will collapse, not the whole system. The typical scale of such a region is what we denote as $\ell$, and it goes down when the pressure increases, in our case like the inverse pressure to the cubic root, see below Eq.~(\ref{ellp}).  Since our screening phenomenon depends on the existence of gradients of the quadrupolar field associated with a plastic avalanche (=region of system collapse), to allow an inverse screening length $\kappa$ (screening length of $\kappa^{-1}$), we need an avalanche of at least that size. This is always available at sufficiently small pressures, but when the pressure is too high the avalanches are too small, not large enough to provide a sufficiently sizeable gradient of the quadrupolar field to form dipoles. There is a critical value of the pressure where the avalanches become large enough to cause a transition between quasi-elastic and anomalous.

\begin{figure} [h]
	\includegraphics[width=1.1\linewidth]{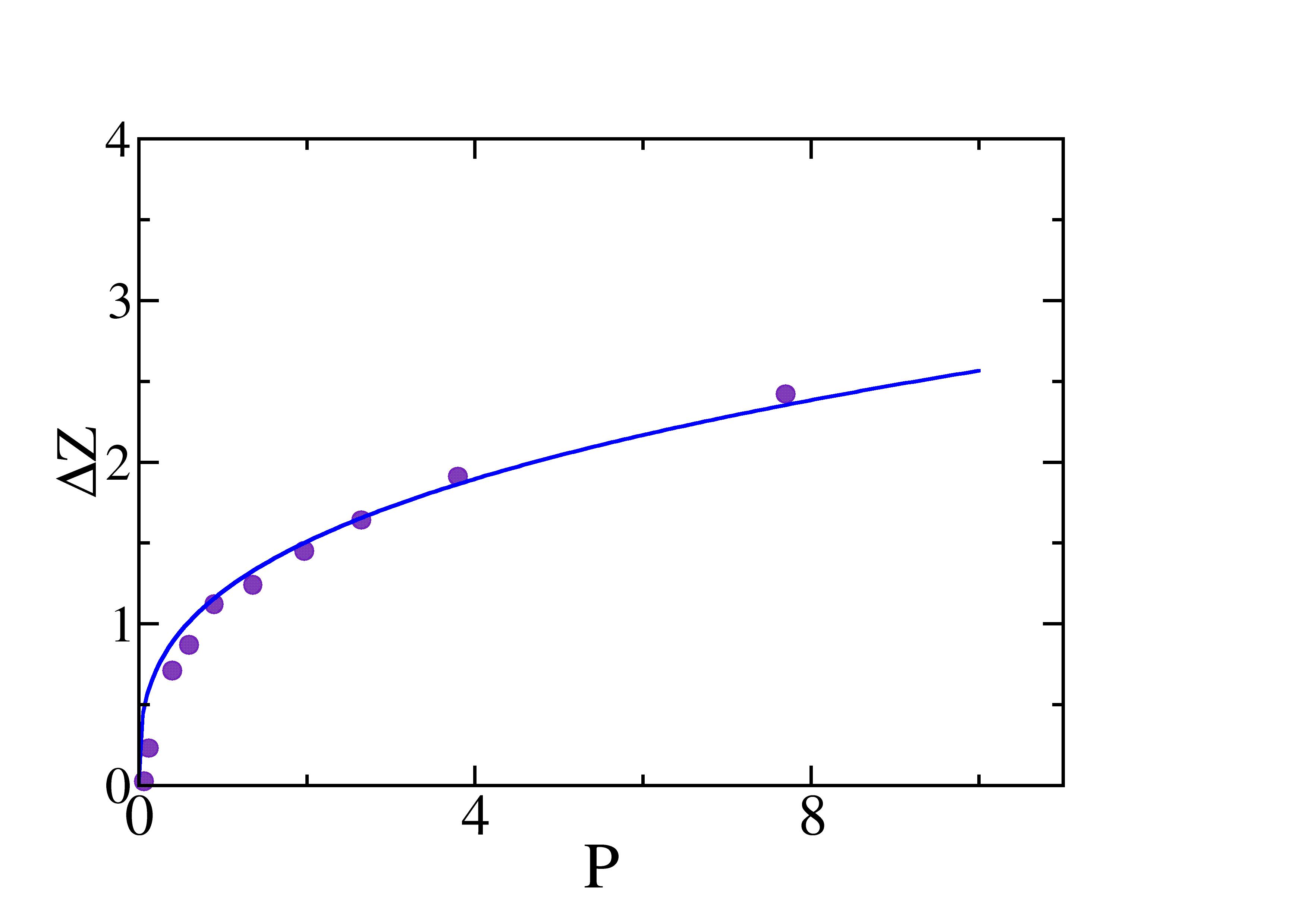}
	\caption{The measured access contacts $\Delta Z$ as a function of the pressure. The continuous line is the power law $\Delta Z \sim 1.15 p^{1/3}$.}
	\label{deltaZ} 
\end{figure}

Start with estimating the maximal size of a blob that can become unstable and go through plastic deformation. In our amorphous configurations of frictionless spheres interacting via Hertzian forces.
the pressure depends on excess coordination number $\Delta Z\equiv Z-Z^*$ according to \cite{10LN,18BMHM}
\begin{equation}
	p\sim (\phi-\phi_J)^{3/2} \sim \Delta Z^3 \ ,
	\label{pvsZ}
\end{equation}
where $Z^*=6$ is the coordination number at jamming. When $\Delta Z=0$, breaking any contact will render the whole system unstable. On the other hand, when $\Delta Z>0$ one can afford breaking more than one bond, in fact, one can break a whole circumference of bonds of length $\ell$ as long as \cite{05WNW},
\begin{equation}
	Z\ell^{d-1} \sim \Delta Z \ell^d \ .
\end{equation}
If the region is larger than this $\ell$, the system would collapse and not remain rigid. We then interpret this length as the maximal blob size that can participate in an avalanche of plastic events. This length scale depends on pressure like
\begin{equation}
	\ell \sim \frac{Z}{\Delta Z} \sim p^{-1/3} \ .
    \label{ellp}
\end{equation}
 Moreover, in Ref.~\cite{09EHS} we find the estimate (cf. Eq.~(21) there), stating that 
$\ell \approx 6/\Delta Z$, or $\Delta Z\approx 6\kappa$. On the other hand, we have measured directly the dependence of $\Delta Z$ on the pressure in our system, see Fig.~\ref{deltaZ}, with the result $\Delta Z\approx 1.15 p^{1/3}$.
Using the average of the measured screening parameter $\kappa\approx 0.256$ (the horizontal line in Fig.~\ref{sharp} panel (a)), we estimate $p_c\approx 2.4$.
This estimate falls right between the estimated value from the jump in panel (a) of Fig.~\ref{sharp} ($p_c\approx 2.3$), and the estimate from the divergent value of $\theta^\dag$, Fig.~\ref{thetadag} (i.e. $p_c\approx 2.9$.) Taking into account
the approximate values of the coefficients in the employed scaling laws of Ref.~\cite{09EHS}, we consider this as a very good agreement. 

We should remark at this point that also the agreement between the predicted value of $\kappa\approx 0.29$ and the measured value of 0.26 is acceptable because the prediction uses a continuum theory whereas the simulation is done with a sphere packing with the outer length $r_{\rm out}$ being only about 31 times larger than our large spheres! Moreover, this agreement is not accidental, changing the inner and outer lengths changes the predicted value of the screening parameter, always in agreement with the corresponding simulations.

{\bf Discussions:} 
The upshot of the analysis is that in correspondence to the Hexatic Transition that showed the existence of an intermediate phase between an ordered solid and a liquid, a phase with broken translation symmetry but with hexatic order, we have here an intermediate phase between an elastic amorphous solid at high pressure and the liquid at zero pressure. The difference is that in our case this intermediate phase exists both in two and three dimensions. The Hexatic Transition deals with the material structure - we discuss the mechanical response of an amorphous solid and the transition is in the nature of the displacement field that results from a strain. In our case, the high-pressure elastic solid responds with a displacement field whose radial component is correlated in both radial and angular directions. In the quasi-elastic regime, all the solutions
of the classical equation for the displacement field are given
by the Michell formalism in terms of periodic functions which
are therefore long-ranged correlated in the angles. The onset
of avalanches destroys these exact results, and the destruction
is more and more severe as the avalanches grow in size and
importance. The liquid responds with a displacement field that is short-ranged in both directions. In the intermediate phase, we lose the angular correlations but maintain the radial long range. The central result of the present paper is that the transition between the intermediate phase and the elastic phase appears critical, in the sense that angular correlations are associated with a diverging correlation length upon approaching the critical pressure $p \approx p_c$. We proposed a scaling argument that succeeds in predicting the transition pressure with satisfactory accuracy. 

We note here that the predicted value of the critical pressure will, in general, depend on the nature of the microscopic interaction law, since $\ell$ depends on the interaction, whereas $\kappa^{-1}$ is determined by the geometry only.

Much remains for future study. For example, how universal is the critical exponent $\mu$ in Eq.~(\ref{result})? Does it depend on the interaction between the particles, would it change for frictional spheres? How does the picture change with different straining protocols? With time-dependent protocols? We hope and trust that these questions would feed interesting research in the near future.


{\bf Acknowledgments}: We thank Yuliang Jin for useful disucssions,   This work has been supported by the joint grant between the Israel Science Foundation and the National Science Foundation of China, and by the Minerva Foundation, Munich, Germany.

\bibliography{3d.critical}

\begin{thebibliography}{36}%
\makeatletter
\providecommand \@ifxundefined [1]{%
 \@ifx{#1\undefined}
}%
\providecommand \@ifnum [1]{%
 \ifnum #1\expandafter \@firstoftwo
 \else \expandafter \@secondoftwo
 \fi
}%
\providecommand \@ifx [1]{%
 \ifx #1\expandafter \@firstoftwo
 \else \expandafter \@secondoftwo
 \fi
}%
\providecommand \natexlab [1]{#1}%
\providecommand \enquote  [1]{``#1''}%
\providecommand \bibnamefont  [1]{#1}%
\providecommand \bibfnamefont [1]{#1}%
\providecommand \citenamefont [1]{#1}%
\providecommand \href@noop [0]{\@secondoftwo}%
\providecommand \href [0]{\begingroup \@sanitize@url \@href}%
\providecommand \@href[1]{\@@startlink{#1}\@@href}%
\providecommand \@@href[1]{\endgroup#1\@@endlink}%
\providecommand \@sanitize@url [0]{\catcode `\\12\catcode `\$12\catcode `\&12\catcode `\#12\catcode `\^12\catcode `\_12\catcode `\%12\relax}%
\providecommand \@@startlink[1]{}%
\providecommand \@@endlink[0]{}%
\providecommand \url  [0]{\begingroup\@sanitize@url \@url }%
\providecommand \@url [1]{\endgroup\@href {#1}{\urlprefix }}%
\providecommand \urlprefix  [0]{URL }%
\providecommand \Eprint [0]{\href }%
\providecommand \doibase [0]{https://doi.org/}%
\providecommand \selectlanguage [0]{\@gobble}%
\providecommand \bibinfo  [0]{\@secondoftwo}%
\providecommand \bibfield  [0]{\@secondoftwo}%
\providecommand \translation [1]{[#1]}%
\providecommand \BibitemOpen [0]{}%
\providecommand \bibitemStop [0]{}%
\providecommand \bibitemNoStop [0]{.\EOS\space}%
\providecommand \EOS [0]{\spacefactor3000\relax}%
\providecommand \BibitemShut  [1]{\csname bibitem#1\endcsname}%
\let\auto@bib@innerbib\@empty
\bibitem [{\citenamefont {Hohenberg}(1967)}]{67Hoh}%
  \BibitemOpen
  \bibfield  {author} {\bibinfo {author} {\bibfnamefont {P.~C.}\ \bibnamefont {Hohenberg}},\ }\bibfield  {title} {\bibinfo {title} {Existence of long-range order in one and two dimensions},\ }\href@noop {} {\bibfield  {journal} {\bibinfo  {journal} {Physical Review}\ }\textbf {\bibinfo {volume} {158}},\ \bibinfo {pages} {383} (\bibinfo {year} {1967})}\BibitemShut {NoStop}%
\bibitem [{\citenamefont {Mermin}\ and\ \citenamefont {Wagner}(1966)}]{66MW}%
  \BibitemOpen
  \bibfield  {author} {\bibinfo {author} {\bibfnamefont {N.~D.}\ \bibnamefont {Mermin}}\ and\ \bibinfo {author} {\bibfnamefont {H.}~\bibnamefont {Wagner}},\ }\bibfield  {title} {\bibinfo {title} {Absence of ferromagnetism or antiferromagnetism in one-or two-dimensional isotropic heisenberg models},\ }\href@noop {} {\bibfield  {journal} {\bibinfo  {journal} {Physical Review Letters}\ }\textbf {\bibinfo {volume} {17}},\ \bibinfo {pages} {1133} (\bibinfo {year} {1966})}\BibitemShut {NoStop}%
\bibitem [{\citenamefont {Kosterlitz}(2016)}]{16Kos}%
  \BibitemOpen
  \bibfield  {author} {\bibinfo {author} {\bibfnamefont {J.~M.}\ \bibnamefont {Kosterlitz}},\ }\bibfield  {title} {\bibinfo {title} {Kosterlitz{\textendash}thouless physics: a review of key issues},\ }\href {https://doi.org/10.1088/0034-4885/79/2/026001} {\bibfield  {journal} {\bibinfo  {journal} {Reports on Progress in Physics}\ }\textbf {\bibinfo {volume} {79}},\ \bibinfo {pages} {026001} (\bibinfo {year} {2016})}\BibitemShut {NoStop}%
\bibitem [{\citenamefont {Kosterlitz}\ and\ \citenamefont {Thouless}(2018)}]{18KT}%
  \BibitemOpen
  \bibfield  {author} {\bibinfo {author} {\bibfnamefont {J.~M.}\ \bibnamefont {Kosterlitz}}\ and\ \bibinfo {author} {\bibfnamefont {D.~J.}\ \bibnamefont {Thouless}},\ }\bibfield  {title} {\bibinfo {title} {Ordering, metastability and phase transitions in two-dimensional systems},\ }in\ \href@noop {} {\emph {\bibinfo {booktitle} {Basic Notions Of Condensed Matter Physics}}}\ (\bibinfo  {publisher} {CRC Press},\ \bibinfo {year} {2018})\ pp.\ \bibinfo {pages} {493--515}\BibitemShut {NoStop}%
\bibitem [{\citenamefont {Nelson}\ and\ \citenamefont {Halperin}(1979)}]{79NH}%
  \BibitemOpen
  \bibfield  {author} {\bibinfo {author} {\bibfnamefont {D.~R.}\ \bibnamefont {Nelson}}\ and\ \bibinfo {author} {\bibfnamefont {B.~I.}\ \bibnamefont {Halperin}},\ }\bibfield  {title} {\bibinfo {title} {Dislocation-mediated melting in two dimensions},\ }\href {https://doi.org/10.1103/PhysRevB.19.2457} {\bibfield  {journal} {\bibinfo  {journal} {Phys. Rev. B}\ }\textbf {\bibinfo {volume} {19}},\ \bibinfo {pages} {2457} (\bibinfo {year} {1979})}\BibitemShut {NoStop}%
\bibitem [{\citenamefont {Lema\^{\i}tre}\ \emph {et~al.}(2021)\citenamefont {Lema\^{\i}tre}, \citenamefont {Mondal}, \citenamefont {Moshe}, \citenamefont {Procaccia}, \citenamefont {Roy},\ and\ \citenamefont {Screiber-Re'em}}]{21LMMPRS}%
  \BibitemOpen
  \bibfield  {author} {\bibinfo {author} {\bibfnamefont {A.}~\bibnamefont {Lema\^{\i}tre}}, \bibinfo {author} {\bibfnamefont {C.}~\bibnamefont {Mondal}}, \bibinfo {author} {\bibfnamefont {M.}~\bibnamefont {Moshe}}, \bibinfo {author} {\bibfnamefont {I.}~\bibnamefont {Procaccia}}, \bibinfo {author} {\bibfnamefont {S.}~\bibnamefont {Roy}},\ and\ \bibinfo {author} {\bibfnamefont {K.}~\bibnamefont {Screiber-Re'em}},\ }\bibfield  {title} {\bibinfo {title} {Anomalous elasticity and plastic screening in amorphous solids},\ }\href {https://doi.org/10.1103/PhysRevE.104.024904} {\bibfield  {journal} {\bibinfo  {journal} {Phys. Rev. E}\ }\textbf {\bibinfo {volume} {104}},\ \bibinfo {pages} {024904} (\bibinfo {year} {2021})}\BibitemShut {NoStop}%
\bibitem [{\citenamefont {Bhowmik}\ \emph {et~al.}(2022)\citenamefont {Bhowmik}, \citenamefont {Moshe},\ and\ \citenamefont {Procaccia}}]{22BMP}%
  \BibitemOpen
  \bibfield  {author} {\bibinfo {author} {\bibfnamefont {B.~P.}\ \bibnamefont {Bhowmik}}, \bibinfo {author} {\bibfnamefont {M.}~\bibnamefont {Moshe}},\ and\ \bibinfo {author} {\bibfnamefont {I.}~\bibnamefont {Procaccia}},\ }\bibfield  {title} {\bibinfo {title} {Direct measurement of dipoles in anomalous elasticity of amorphous solids},\ }\href {https://doi.org/10.1103/PhysRevE.105.L043001} {\bibfield  {journal} {\bibinfo  {journal} {Phys. Rev. E}\ }\textbf {\bibinfo {volume} {105}},\ \bibinfo {pages} {L043001} (\bibinfo {year} {2022})}\BibitemShut {NoStop}%
\bibitem [{\citenamefont {Charan}\ \emph {et~al.}(2023)\citenamefont {Charan}, \citenamefont {Moshe},\ and\ \citenamefont {Procaccia}}]{23CMP}%
  \BibitemOpen
  \bibfield  {author} {\bibinfo {author} {\bibfnamefont {H.}~\bibnamefont {Charan}}, \bibinfo {author} {\bibfnamefont {M.}~\bibnamefont {Moshe}},\ and\ \bibinfo {author} {\bibfnamefont {I.}~\bibnamefont {Procaccia}},\ }\bibfield  {title} {\bibinfo {title} {Anomalous elasticity and emergent dipole screening in three-dimensional amorphous solids},\ }\href {https://doi.org/10.1103/PhysRevE.107.055005} {\bibfield  {journal} {\bibinfo  {journal} {Phys. Rev. E}\ }\textbf {\bibinfo {volume} {107}},\ \bibinfo {pages} {055005} (\bibinfo {year} {2023})}\BibitemShut {NoStop}%
\bibitem [{\citenamefont {Mondal}\ \emph {et~al.}(2022)\citenamefont {Mondal}, \citenamefont {Moshe}, \citenamefont {Procaccia}, \citenamefont {Roy}, \citenamefont {Shang},\ and\ \citenamefont {Zhang}}]{22MMPRSZ}%
  \BibitemOpen
  \bibfield  {author} {\bibinfo {author} {\bibfnamefont {C.}~\bibnamefont {Mondal}}, \bibinfo {author} {\bibfnamefont {M.}~\bibnamefont {Moshe}}, \bibinfo {author} {\bibfnamefont {I.}~\bibnamefont {Procaccia}}, \bibinfo {author} {\bibfnamefont {S.}~\bibnamefont {Roy}}, \bibinfo {author} {\bibfnamefont {J.}~\bibnamefont {Shang}},\ and\ \bibinfo {author} {\bibfnamefont {J.}~\bibnamefont {Zhang}},\ }\bibfield  {title} {\bibinfo {title} {Experimental and numerical verification of anomalous screening theory in granular matter},\ }\href {https://doi.org/https://doi.org/10.1016/j.chaos.2022.112609} {\bibfield  {journal} {\bibinfo  {journal} {Chaos, Solitons and Fractals}\ }\textbf {\bibinfo {volume} {164}},\ \bibinfo {pages} {112609} (\bibinfo {year} {2022})}\BibitemShut {NoStop}%
\bibitem [{\citenamefont {Kumar}\ \emph {et~al.}(2022)\citenamefont {Kumar}, \citenamefont {Moshe}, \citenamefont {Procaccia},\ and\ \citenamefont {Singh}}]{22KMPS}%
  \BibitemOpen
  \bibfield  {author} {\bibinfo {author} {\bibfnamefont {A.}~\bibnamefont {Kumar}}, \bibinfo {author} {\bibfnamefont {M.}~\bibnamefont {Moshe}}, \bibinfo {author} {\bibfnamefont {I.}~\bibnamefont {Procaccia}},\ and\ \bibinfo {author} {\bibfnamefont {M.}~\bibnamefont {Singh}},\ }\bibfield  {title} {\bibinfo {title} {Anomalous elasticity in classical glass formers},\ }\href {https://doi.org/10.1103/PhysRevE.106.015001} {\bibfield  {journal} {\bibinfo  {journal} {Phys. Rev. E}\ }\textbf {\bibinfo {volume} {106}},\ \bibinfo {pages} {015001} (\bibinfo {year} {2022})}\BibitemShut {NoStop}%
\bibitem [{\citenamefont {Mondal}\ \emph {et~al.}(2023)\citenamefont {Mondal}, \citenamefont {Moshe}, \citenamefont {Procaccia},\ and\ \citenamefont {Roy}}]{23MMPR}%
  \BibitemOpen
  \bibfield  {author} {\bibinfo {author} {\bibfnamefont {C.}~\bibnamefont {Mondal}}, \bibinfo {author} {\bibfnamefont {M.}~\bibnamefont {Moshe}}, \bibinfo {author} {\bibfnamefont {I.}~\bibnamefont {Procaccia}},\ and\ \bibinfo {author} {\bibfnamefont {S.}~\bibnamefont {Roy}},\ }\href@noop {} {\bibinfo {title} {Dipole screening in pure shear strain protocols of amorphous solids}} (\bibinfo {year} {2023}),\ \Eprint {https://arxiv.org/abs/2305.11253} {arXiv:2305.11253 [cond-mat.soft Phys.Rev. E, in press]} \BibitemShut {NoStop}%
\bibitem [{\citenamefont {Hentschel}\ \emph {et~al.}(2024)\citenamefont {Hentschel}, \citenamefont {Pomyalov}, \citenamefont {Procaccia},\ and\ \citenamefont {Szachter}}]{24HPPS}%
  \BibitemOpen
  \bibfield  {author} {\bibinfo {author} {\bibfnamefont {H.~G.~E.}\ \bibnamefont {Hentschel}}, \bibinfo {author} {\bibfnamefont {A.}~\bibnamefont {Pomyalov}}, \bibinfo {author} {\bibfnamefont {I.}~\bibnamefont {Procaccia}},\ and\ \bibinfo {author} {\bibfnamefont {O.}~\bibnamefont {Szachter}},\ }\bibfield  {title} {\bibinfo {title} {Dynamic screening by plasticity in amorphous solids},\ }\href {https://doi.org/10.1103/PhysRevE.109.044902} {\bibfield  {journal} {\bibinfo  {journal} {Phys. Rev. E}\ }\textbf {\bibinfo {volume} {109}},\ \bibinfo {pages} {044902} (\bibinfo {year} {2024})}\BibitemShut {NoStop}%
\bibitem [{\citenamefont {Karmakar}\ \emph {et~al.}(2010)\citenamefont {Karmakar}, \citenamefont {Lerner},\ and\ \citenamefont {Procaccia}}]{10KLP}%
  \BibitemOpen
  \bibfield  {author} {\bibinfo {author} {\bibfnamefont {S.}~\bibnamefont {Karmakar}}, \bibinfo {author} {\bibfnamefont {E.}~\bibnamefont {Lerner}},\ and\ \bibinfo {author} {\bibfnamefont {I.}~\bibnamefont {Procaccia}},\ }\bibfield  {title} {\bibinfo {title} {{Athermal nonlinear elastic constants of amorphous solids}},\ }\href {https://doi.org/10.1103/PhysRevE.82.026105} {\bibfield  {journal} {\bibinfo  {journal} {Phys. Rev.E}\ }\textbf {\bibinfo {volume} {82}},\ \bibinfo {pages} {026105} (\bibinfo {year} {2010})}\BibitemShut {NoStop}%
\bibitem [{\citenamefont {Hentschel}\ \emph {et~al.}(2011)\citenamefont {Hentschel}, \citenamefont {Karmakar}, \citenamefont {Lerner},\ and\ \citenamefont {Procaccia}}]{11HKLP}%
  \BibitemOpen
  \bibfield  {author} {\bibinfo {author} {\bibfnamefont {H.~G.~E.}\ \bibnamefont {Hentschel}}, \bibinfo {author} {\bibfnamefont {S.}~\bibnamefont {Karmakar}}, \bibinfo {author} {\bibfnamefont {E.}~\bibnamefont {Lerner}},\ and\ \bibinfo {author} {\bibfnamefont {I.}~\bibnamefont {Procaccia}},\ }\bibfield  {title} {\bibinfo {title} {{Do athermal amorphous solids exist?}},\ }\href {https://doi.org/10.1103/PhysRevE.83.061101} {\bibfield  {journal} {\bibinfo  {journal} {Phys. Rev.E}\ }\textbf {\bibinfo {volume} {83}},\ \bibinfo {pages} {061101} (\bibinfo {year} {2011})}\BibitemShut {NoStop}%
\bibitem [{\citenamefont {Eshelby}(1957)}]{54Esh}%
  \BibitemOpen
  \bibfield  {author} {\bibinfo {author} {\bibfnamefont {J.~D.}\ \bibnamefont {Eshelby}},\ }\bibfield  {title} {\bibinfo {title} {The determination of the elastic field of an ellipsoidal inclusion, and related problems},\ }\href {https://doi.org/10.1098/rspa.1957.0133} {\bibfield  {journal} {\bibinfo  {journal} {Proceedings of the Royal Society of London A: Mathematical, Physical and Engineering Sciences}\ }\textbf {\bibinfo {volume} {241}},\ \bibinfo {pages} {376} (\bibinfo {year} {1957})}\BibitemShut {NoStop}%
\bibitem [{\citenamefont {Cohen}\ \emph {et~al.}(2024)\citenamefont {Cohen}, \citenamefont {Schiller}, \citenamefont {Wang}, \citenamefont {Dijksman},\ and\ \citenamefont {Moshe}}]{MM24odd}%
  \BibitemOpen
  \bibfield  {author} {\bibinfo {author} {\bibfnamefont {Y.}~\bibnamefont {Cohen}}, \bibinfo {author} {\bibfnamefont {A.}~\bibnamefont {Schiller}}, \bibinfo {author} {\bibfnamefont {D.}~\bibnamefont {Wang}}, \bibinfo {author} {\bibfnamefont {J.}~\bibnamefont {Dijksman}},\ and\ \bibinfo {author} {\bibfnamefont {M.}~\bibnamefont {Moshe}},\ }\href {https://arxiv.org/abs/2310.09942} {\bibinfo {title} {Odd dipole screening in disordered matter}} (\bibinfo {year} {2024}),\ \Eprint {https://arxiv.org/abs/2310.09942} {arXiv:2310.09942 [cond-mat.soft]} \BibitemShut {NoStop}%
\bibitem [{\citenamefont {Kaur}\ \emph {et~al.}(2025)\citenamefont {Kaur}, \citenamefont {Procaccia},\ and\ \citenamefont {Samanta}}]{TS25}%
  \BibitemOpen
  \bibfield  {author} {\bibinfo {author} {\bibfnamefont {P.}~\bibnamefont {Kaur}}, \bibinfo {author} {\bibfnamefont {I.}~\bibnamefont {Procaccia}},\ and\ \bibinfo {author} {\bibfnamefont {T.}~\bibnamefont {Samanta}},\ }\bibfield  {title} {\bibinfo {title} {Selection principle for the screening parameters in the mechanical response of amorphous solids},\ }\href {https://doi.org/10.1103/PhysRevE.111.015506} {\bibfield  {journal} {\bibinfo  {journal} {Phys. Rev. E}\ }\textbf {\bibinfo {volume} {111}},\ \bibinfo {pages} {015506} (\bibinfo {year} {2025})}\BibitemShut {NoStop}%
\bibitem [{\citenamefont {Livne}\ \emph {et~al.}(2024)\citenamefont {Livne}, \citenamefont {Samanta}, \citenamefont {Schiller}, \citenamefont {Procaccia},\ and\ \citenamefont {Moshe}}]{24LSSPM}%
  \BibitemOpen
  \bibfield  {author} {\bibinfo {author} {\bibfnamefont {N.~S.}\ \bibnamefont {Livne}}, \bibinfo {author} {\bibfnamefont {T.}~\bibnamefont {Samanta}}, \bibinfo {author} {\bibfnamefont {A.}~\bibnamefont {Schiller}}, \bibinfo {author} {\bibfnamefont {I.}~\bibnamefont {Procaccia}},\ and\ \bibinfo {author} {\bibfnamefont {M.}~\bibnamefont {Moshe}},\ }\href {https://arxiv.org/abs/2408.13086} {\bibinfo {title} {Continuum mechanics of differential growth in disordered granular matter}} (\bibinfo {year} {2024}),\ \Eprint {https://arxiv.org/abs/2408.13086} {arXiv:2408.13086 [cond-mat.soft]} \BibitemShut {NoStop}%
\bibitem [{\citenamefont {Baule}\ \emph {et~al.}(2018)\citenamefont {Baule}, \citenamefont {Morone}, \citenamefont {Herrmann},\ and\ \citenamefont {Makse}}]{18BMHM}%
  \BibitemOpen
  \bibfield  {author} {\bibinfo {author} {\bibfnamefont {A.}~\bibnamefont {Baule}}, \bibinfo {author} {\bibfnamefont {F.}~\bibnamefont {Morone}}, \bibinfo {author} {\bibfnamefont {H.~J.}\ \bibnamefont {Herrmann}},\ and\ \bibinfo {author} {\bibfnamefont {H.~A.}\ \bibnamefont {Makse}},\ }\bibfield  {title} {\bibinfo {title} {Edwards statistical mechanics for jammed granular matter},\ }\href {https://doi.org/10.1103/RevModPhys.90.015006} {\bibfield  {journal} {\bibinfo  {journal} {Rev. Mod. Phys.}\ }\textbf {\bibinfo {volume} {90}},\ \bibinfo {pages} {015006} (\bibinfo {year} {2018})}\BibitemShut {NoStop}%
\bibitem [{\citenamefont {O'Hern}\ \emph {et~al.}(2003)\citenamefont {O'Hern}, \citenamefont {Silbert}, \citenamefont {Liu},\ and\ \citenamefont {Nagel}}]{OH03}%
  \BibitemOpen
  \bibfield  {author} {\bibinfo {author} {\bibfnamefont {C.~S.}\ \bibnamefont {O'Hern}}, \bibinfo {author} {\bibfnamefont {L.~E.}\ \bibnamefont {Silbert}}, \bibinfo {author} {\bibfnamefont {A.~J.}\ \bibnamefont {Liu}},\ and\ \bibinfo {author} {\bibfnamefont {S.~R.}\ \bibnamefont {Nagel}},\ }\bibfield  {title} {\bibinfo {title} {Jamming at zero temperature and zero applied stress: The epitome of disorder},\ }\href {https://doi.org/10.1103/PhysRevE.68.011306} {\bibfield  {journal} {\bibinfo  {journal} {Phys. Rev. E}\ }\textbf {\bibinfo {volume} {68}},\ \bibinfo {pages} {011306} (\bibinfo {year} {2003})}\BibitemShut {NoStop}%
\bibitem [{\citenamefont {Makse}\ \emph {et~al.}(2000)\citenamefont {Makse}, \citenamefont {Johnson},\ and\ \citenamefont {Schwartz}}]{ML00}%
  \BibitemOpen
  \bibfield  {author} {\bibinfo {author} {\bibfnamefont {H.~A.}\ \bibnamefont {Makse}}, \bibinfo {author} {\bibfnamefont {D.~L.}\ \bibnamefont {Johnson}},\ and\ \bibinfo {author} {\bibfnamefont {L.~M.}\ \bibnamefont {Schwartz}},\ }\bibfield  {title} {\bibinfo {title} {Packing of compressible granular materials},\ }\href {https://doi.org/10.1103/PhysRevLett.84.4160} {\bibfield  {journal} {\bibinfo  {journal} {Phys. Rev. Lett.}\ }\textbf {\bibinfo {volume} {84}},\ \bibinfo {pages} {4160} (\bibinfo {year} {2000})}\BibitemShut {NoStop}%
\bibitem [{\citenamefont {Chaudhuri}\ \emph {et~al.}(2010)\citenamefont {Chaudhuri}, \citenamefont {Berthier},\ and\ \citenamefont {Sastry}}]{10CBS}%
  \BibitemOpen
  \bibfield  {author} {\bibinfo {author} {\bibfnamefont {P.}~\bibnamefont {Chaudhuri}}, \bibinfo {author} {\bibfnamefont {L.}~\bibnamefont {Berthier}},\ and\ \bibinfo {author} {\bibfnamefont {S.}~\bibnamefont {Sastry}},\ }\bibfield  {title} {\bibinfo {title} {Jamming transitions in amorphous packings of frictionless spheres occur over a continuous range of volume fractions},\ }\href {https://doi.org/10.1103/PhysRevLett.104.165701} {\bibfield  {journal} {\bibinfo  {journal} {Phys. Rev. Lett.}\ }\textbf {\bibinfo {volume} {104}},\ \bibinfo {pages} {165701} (\bibinfo {year} {2010})}\BibitemShut {NoStop}%
\bibitem [{\citenamefont {Charbonneau}\ \emph {et~al.}(2012)\citenamefont {Charbonneau}, \citenamefont {Corwin}, \citenamefont {Parisi},\ and\ \citenamefont {Zamponi}}]{FZ12}%
  \BibitemOpen
  \bibfield  {author} {\bibinfo {author} {\bibfnamefont {P.}~\bibnamefont {Charbonneau}}, \bibinfo {author} {\bibfnamefont {E.~I.}\ \bibnamefont {Corwin}}, \bibinfo {author} {\bibfnamefont {G.}~\bibnamefont {Parisi}},\ and\ \bibinfo {author} {\bibfnamefont {F.}~\bibnamefont {Zamponi}},\ }\bibfield  {title} {\bibinfo {title} {Universal microstructure and mechanical stability of jammed packings},\ }\href {https://doi.org/10.1103/PhysRevLett.109.205501} {\bibfield  {journal} {\bibinfo  {journal} {Phys. Rev. Lett.}\ }\textbf {\bibinfo {volume} {109}},\ \bibinfo {pages} {205501} (\bibinfo {year} {2012})}\BibitemShut {NoStop}%
\bibitem [{\citenamefont {Morse}\ and\ \citenamefont {Corwin}(2014)}]{MV14}%
  \BibitemOpen
  \bibfield  {author} {\bibinfo {author} {\bibfnamefont {P.~K.}\ \bibnamefont {Morse}}\ and\ \bibinfo {author} {\bibfnamefont {E.~I.}\ \bibnamefont {Corwin}},\ }\bibfield  {title} {\bibinfo {title} {Geometric signatures of jamming in the mechanical vacuum},\ }\href {https://doi.org/10.1103/PhysRevLett.112.115701} {\bibfield  {journal} {\bibinfo  {journal} {Phys. Rev. Lett.}\ }\textbf {\bibinfo {volume} {112}},\ \bibinfo {pages} {115701} (\bibinfo {year} {2014})}\BibitemShut {NoStop}%
\bibitem [{\citenamefont {O'Hern}\ \emph {et~al.}(2002)\citenamefont {O'Hern}, \citenamefont {Langer}, \citenamefont {Liu},\ and\ \citenamefont {Nagel}}]{NL02}%
  \BibitemOpen
  \bibfield  {author} {\bibinfo {author} {\bibfnamefont {C.~S.}\ \bibnamefont {O'Hern}}, \bibinfo {author} {\bibfnamefont {S.~A.}\ \bibnamefont {Langer}}, \bibinfo {author} {\bibfnamefont {A.~J.}\ \bibnamefont {Liu}},\ and\ \bibinfo {author} {\bibfnamefont {S.~R.}\ \bibnamefont {Nagel}},\ }\bibfield  {title} {\bibinfo {title} {Random packings of frictionless particles},\ }\href {https://doi.org/10.1103/PhysRevLett.88.075507} {\bibfield  {journal} {\bibinfo  {journal} {Phys. Rev. Lett.}\ }\textbf {\bibinfo {volume} {88}},\ \bibinfo {pages} {075507} (\bibinfo {year} {2002})}\BibitemShut {NoStop}%
\bibitem [{\citenamefont {Heussinger}\ and\ \citenamefont {Barrat}(2009)}]{JL09}%
  \BibitemOpen
  \bibfield  {author} {\bibinfo {author} {\bibfnamefont {C.}~\bibnamefont {Heussinger}}\ and\ \bibinfo {author} {\bibfnamefont {J.-L.}\ \bibnamefont {Barrat}},\ }\bibfield  {title} {\bibinfo {title} {Jamming transition as probed by quasistatic shear flow},\ }\href {https://doi.org/10.1103/PhysRevLett.102.218303} {\bibfield  {journal} {\bibinfo  {journal} {Phys. Rev. Lett.}\ }\textbf {\bibinfo {volume} {102}},\ \bibinfo {pages} {218303} (\bibinfo {year} {2009})}\BibitemShut {NoStop}%
\bibitem [{\citenamefont {Pan}\ \emph {et~al.}(2023)\citenamefont {Pan}, \citenamefont {Wang}, \citenamefont {Yoshino}, \citenamefont {Zhang},\ and\ \citenamefont {Jin}}]{PAN2023}%
  \BibitemOpen
  \bibfield  {author} {\bibinfo {author} {\bibfnamefont {D.}~\bibnamefont {Pan}}, \bibinfo {author} {\bibfnamefont {Y.}~\bibnamefont {Wang}}, \bibinfo {author} {\bibfnamefont {H.}~\bibnamefont {Yoshino}}, \bibinfo {author} {\bibfnamefont {J.}~\bibnamefont {Zhang}},\ and\ \bibinfo {author} {\bibfnamefont {Y.}~\bibnamefont {Jin}},\ }\bibfield  {title} {\bibinfo {title} {A review on shear jamming},\ }\href {https://doi.org/https://doi.org/10.1016/j.physrep.2023.10.002} {\bibfield  {journal} {\bibinfo  {journal} {Physics Reports}\ }\textbf {\bibinfo {volume} {1038}},\ \bibinfo {pages} {1} (\bibinfo {year} {2023})},\ \bibinfo {note} {a review on shear jamming}\BibitemShut {NoStop}%
\bibitem [{\citenamefont {V\aa{}gberg}\ \emph {et~al.}(2011)\citenamefont {V\aa{}gberg}, \citenamefont {Valdez-Balderas}, \citenamefont {Moore}, \citenamefont {Olsson},\ and\ \citenamefont {Teitel}}]{PT11}%
  \BibitemOpen
  \bibfield  {author} {\bibinfo {author} {\bibfnamefont {D.}~\bibnamefont {V\aa{}gberg}}, \bibinfo {author} {\bibfnamefont {D.}~\bibnamefont {Valdez-Balderas}}, \bibinfo {author} {\bibfnamefont {M.~A.}\ \bibnamefont {Moore}}, \bibinfo {author} {\bibfnamefont {P.}~\bibnamefont {Olsson}},\ and\ \bibinfo {author} {\bibfnamefont {S.}~\bibnamefont {Teitel}},\ }\bibfield  {title} {\bibinfo {title} {Finite-size scaling at the jamming transition: Corrections to scaling and the correlation-length critical exponent},\ }\href {https://doi.org/10.1103/PhysRevE.83.030303} {\bibfield  {journal} {\bibinfo  {journal} {Phys. Rev. E}\ }\textbf {\bibinfo {volume} {83}},\ \bibinfo {pages} {030303} (\bibinfo {year} {2011})}\BibitemShut {NoStop}%
\bibitem [{\citenamefont {Thompson}\ and\ \citenamefont {Clark}(2019)}]{Abram19}%
  \BibitemOpen
  \bibfield  {author} {\bibinfo {author} {\bibfnamefont {J.~D.}\ \bibnamefont {Thompson}}\ and\ \bibinfo {author} {\bibfnamefont {A.~H.}\ \bibnamefont {Clark}},\ }\bibfield  {title} {\bibinfo {title} {Critical scaling for yield is independent of distance to isostaticity},\ }\href {https://doi.org/10.1103/PhysRevResearch.1.012002} {\bibfield  {journal} {\bibinfo  {journal} {Phys. Rev. Res.}\ }\textbf {\bibinfo {volume} {1}},\ \bibinfo {pages} {012002} (\bibinfo {year} {2019})}\BibitemShut {NoStop}%
\bibitem [{\citenamefont {Clark}\ \emph {et~al.}(2018)\citenamefont {Clark}, \citenamefont {Thompson}, \citenamefont {Shattuck}, \citenamefont {Ouellette},\ and\ \citenamefont {O'Hern}}]{HN18}%
  \BibitemOpen
  \bibfield  {author} {\bibinfo {author} {\bibfnamefont {A.~H.}\ \bibnamefont {Clark}}, \bibinfo {author} {\bibfnamefont {J.~D.}\ \bibnamefont {Thompson}}, \bibinfo {author} {\bibfnamefont {M.~D.}\ \bibnamefont {Shattuck}}, \bibinfo {author} {\bibfnamefont {N.~T.}\ \bibnamefont {Ouellette}},\ and\ \bibinfo {author} {\bibfnamefont {C.~S.}\ \bibnamefont {O'Hern}},\ }\bibfield  {title} {\bibinfo {title} {Critical scaling near the yielding transition in granular media},\ }\href {https://doi.org/10.1103/PhysRevE.97.062901} {\bibfield  {journal} {\bibinfo  {journal} {Phys. Rev. E}\ }\textbf {\bibinfo {volume} {97}},\ \bibinfo {pages} {062901} (\bibinfo {year} {2018})}\BibitemShut {NoStop}%
\bibitem [{\citenamefont {Fu}\ \emph {et~al.}(2024)\citenamefont {Fu}, \citenamefont {Jin}, \citenamefont {Pan},\ and\ \citenamefont {Procaccia}}]{fu24}%
  \BibitemOpen
  \bibfield  {author} {\bibinfo {author} {\bibfnamefont {Y.}~\bibnamefont {Fu}}, \bibinfo {author} {\bibfnamefont {Y.}~\bibnamefont {Jin}}, \bibinfo {author} {\bibfnamefont {D.}~\bibnamefont {Pan}},\ and\ \bibinfo {author} {\bibfnamefont {I.}~\bibnamefont {Procaccia}},\ }\href {https://arxiv.org/abs/2410.04138} {\bibinfo {title} {Long-range angular correlations of particle displacements at a plastic-to-elastic transition in jammed amorphous solids}} (\bibinfo {year} {2024}),\ \Eprint {https://arxiv.org/abs/2410.04138} {arXiv:2410.04138 [cond-mat.soft]} \BibitemShut {NoStop}%
\bibitem [{\citenamefont {Jin}\ \emph {et~al.}(2024)\citenamefont {Jin}, \citenamefont {Procaccia},\ and\ \citenamefont {Samanta}}]{24JPS}%
  \BibitemOpen
  \bibfield  {author} {\bibinfo {author} {\bibfnamefont {Y.}~\bibnamefont {Jin}}, \bibinfo {author} {\bibfnamefont {I.}~\bibnamefont {Procaccia}},\ and\ \bibinfo {author} {\bibfnamefont {T.}~\bibnamefont {Samanta}},\ }\bibfield  {title} {\bibinfo {title} {Intermediate phase between jammed and unjammed amorphous solids},\ }\href {https://doi.org/10.1103/PhysRevE.109.014902} {\bibfield  {journal} {\bibinfo  {journal} {Phys. Rev. E}\ }\textbf {\bibinfo {volume} {109}},\ \bibinfo {pages} {014902} (\bibinfo {year} {2024})}\BibitemShut {NoStop}%
\bibitem [{\citenamefont {Toda}\ \emph {et~al.}(2012)\citenamefont {Toda}, \citenamefont {Kubo}, \citenamefont {Kubo}, \citenamefont {Toda}, \citenamefont {Saito}, \citenamefont {Hashitsume},\ and\ \citenamefont {Hashitsume}}]{toda2012}%
  \BibitemOpen
  \bibfield  {author} {\bibinfo {author} {\bibfnamefont {M.}~\bibnamefont {Toda}}, \bibinfo {author} {\bibfnamefont {R.}~\bibnamefont {Kubo}}, \bibinfo {author} {\bibfnamefont {R.}~\bibnamefont {Kubo}}, \bibinfo {author} {\bibfnamefont {M.}~\bibnamefont {Toda}}, \bibinfo {author} {\bibfnamefont {N.}~\bibnamefont {Saito}}, \bibinfo {author} {\bibfnamefont {N.}~\bibnamefont {Hashitsume}},\ and\ \bibinfo {author} {\bibfnamefont {N.}~\bibnamefont {Hashitsume}},\ }\href {https://books.google.co.il/books?id=cF3wCAAAQBAJ} {\emph {\bibinfo {title} {Statistical Physics II: Nonequilibrium Statistical Mechanics}}},\ Springer Series in Solid-State Sciences\ (\bibinfo  {publisher} {Springer Berlin Heidelberg},\ \bibinfo {year} {2012})\BibitemShut {NoStop}%
\bibitem [{\citenamefont {Liu}\ and\ \citenamefont {Nagel}(2010)}]{10LN}%
  \BibitemOpen
  \bibfield  {author} {\bibinfo {author} {\bibfnamefont {A.~J.}\ \bibnamefont {Liu}}\ and\ \bibinfo {author} {\bibfnamefont {S.~R.}\ \bibnamefont {Nagel}},\ }\bibfield  {title} {\bibinfo {title} {The jamming transition and the marginally jammed solid},\ }\href {https://doi.org/10.1146/annurev-conmatphys-070909-104045} {\bibfield  {journal} {\bibinfo  {journal} {Annual Review of Condensed Matter Physics}\ }\textbf {\bibinfo {volume} {1}},\ \bibinfo {pages} {347} (\bibinfo {year} {2010})}\BibitemShut {NoStop}%
\bibitem [{\citenamefont {Wyart}\ \emph {et~al.}(2005)\citenamefont {Wyart}, \citenamefont {Nagel},\ and\ \citenamefont {Witten}}]{05WNW}%
  \BibitemOpen
  \bibfield  {author} {\bibinfo {author} {\bibfnamefont {M.}~\bibnamefont {Wyart}}, \bibinfo {author} {\bibfnamefont {S.~R.}\ \bibnamefont {Nagel}},\ and\ \bibinfo {author} {\bibfnamefont {T.~A.}\ \bibnamefont {Witten}},\ }\bibfield  {title} {\bibinfo {title} {Geometric origin of excess low-frequency vibrational modes in weakly connected amorphous solids},\ }\href {https://doi.org/10.1209/epl/i2005-10245-5} {\bibfield  {journal} {\bibinfo  {journal} {Europhysics Letters}\ }\textbf {\bibinfo {volume} {72}},\ \bibinfo {pages} {486} (\bibinfo {year} {2005})}\BibitemShut {NoStop}%
\bibitem [{\citenamefont {Ellenbroek}\ \emph {et~al.}(2009)\citenamefont {Ellenbroek}, \citenamefont {van Hecke},\ and\ \citenamefont {van Saarloos}}]{09EHS}%
  \BibitemOpen
  \bibfield  {author} {\bibinfo {author} {\bibfnamefont {W.~G.}\ \bibnamefont {Ellenbroek}}, \bibinfo {author} {\bibfnamefont {M.}~\bibnamefont {van Hecke}},\ and\ \bibinfo {author} {\bibfnamefont {W.}~\bibnamefont {van Saarloos}},\ }\bibfield  {title} {\bibinfo {title} {Jammed frictionless disks: Connecting local and global response},\ }\href {https://doi.org/10.1103/PhysRevE.80.061307} {\bibfield  {journal} {\bibinfo  {journal} {Phys. Rev. E}\ }\textbf {\bibinfo {volume} {80}},\ \bibinfo {pages} {061307} (\bibinfo {year} {2009})}\BibitemShut {NoStop}%
\end{thebibliography}%

\end{document}